\documentclass[12pt]{iopart}
\usepackage{iopams}
\usepackage{graphicx}
\usepackage[nottoc]{tocbibind}
\usepackage{color}
\usepackage{booktabs}
\usepackage{textcomp}
\usepackage{subcaption}
\usepackage{float}
\usepackage{natbib}
\usepackage{soul}
\usepackage[colorlinks=true, linkcolor=blue, citecolor=blue]{hyperref}
\makeatletter
\providecommand{\newblock}{}
\makeatother

\begin{document}

\title[Cu-based Nb$_3$Sn Preparation via Bronze Route]
{Preparation of the First Cu-based Nb$_3$Sn Sample via Bronze Route for Quadrupole Resonator Testing}

\author{Ming Lu$^{1,2,*}$, Sebastian Keckert$^1$, Felix Kramer$^{1}$, Alena Prudnikava$^1$, Jens Knobloch$^{1,3}$, Aleksandr Zubtsovskii$^3$ and Oliver Kugeler$^{1}$}
\address{$^1$ Helmholtz-Zentrum Berlin für Materialien und Energie GmbH, 14109 Berlin, Germany\\
$^2$ Institute of Modern Physics, Chinese Academy of Sciences, Lanzhou, 730000, China\\
$^3$ Universität Siegen, 57076 Siegen, Germany}
\ead{ming.lu@helmholtz-berlin.de}
\vspace{10pt}
\begin{indented}
\item[] $^*$Author to whom any correspondence should be addressed.
\end{indented}

\begin{abstract}
We report the first successful production of a Cu-based Nb$_3$Sn sample specifically designed for Quadrupole Resonator (QPR) testing, representing a significant step toward scalable RF superconducting coatings of Nb$_3$Sn on copper substrates. The sample was fabricated using an optimized electrochemical thermal synthesis (ETS) via the bronze route, incorporating several key advancements: electropolishing of the Cu substrate, electroplating of the bronze precursor layer, a tailored heat treatment at approximately 700\,$^\circ$C to promote grain growth and suppress tin-rich impurity phases, and a newly developed chemical etching procedure for effective removal of surface bronze residues and contaminants. These improvements address longstanding challenges in the fabrication of high-quality Cu-based Nb$_3$Sn thin films. Subsequent QPR measurements yielded the peak magnetic field and temperature dependent surface resistance $R\textsubscript{s}$, as well as the superconducting transition temperature and quench field. Although the achieved RF performance---characterized by a minimum $R\textsubscript{s}$ of 43.4\,n$\Omega$ at 4.5\,K and 15\,mT---is not yet optimal, the results clearly demonstrate the feasibility of this approach and its potential for further enhancement through process refinement.
\end{abstract}

\vspace{2pc}

\section{Introduction}

Nb$_3$Sn has emerged as a promising material for next-generation superconducting radio-frequency (SRF) cavities due to its high critical temperature ($T\textsubscript{c} \approx 18$\,K), low BCS surface resistance, and efficient operation at 4.2\,K, which enables simplified and cost-effective cryogenic systems~\cite{Posen:APL2015,becker2015}. Significant progress has been made using tin vapor diffusion to coat Nb substrates with Nb$_3$Sn, yielding excellent RF performance~\cite{JIANG2024,Posen_2021,Lee_2019,Posen_2017}. However, extending Nb$_3$Sn coatings to copper (Cu) substrates—an attractive alternative due to Cu's superior thermal conductivity, mechanical properties, and lower cost—remains a considerable challenge~\cite{Ilyina2019}.

Cu-based SRF cavities offer compelling advantages, particularly through their compatibility with cryocooler-based conduction cooling
\cite{Stilin_2023,Ciovati_SUST2020}. The high thermal conductivity of Cu facilitates efficient heat extraction, potentially eliminating the need for liquid helium and enabling operation at elevated temperatures (4.2–4.5\,K). Such configurations are highly desirable for compact accelerators in industrial, medical, and space environments~\cite{E.barzi2022}. Despite these benefits, synthesizing high-quality, low-loss Nb$_3$Sn films on Cu substrates has been hindered by several issues, including a loose and porous microstructure of the Nb$_3$Sn layer, poor interfacial adhesion, film cracking induced by thermal expansion mismatch, and the formation of non-stoichiometric or impurity phases~\cite{Ilyina2019,Barzi2021}.

Various thin-film fabrication techniques have been explored for Cu-based Nb$_3$Sn, such as tin vapor diffusion combined with outer-wall copper coating~\cite{Ciovati2023}, magnetron sputtering~\cite{Ilyina2019,SAYEED2021}, chemical vapor deposition (CVD)~\cite{Carta:TFSRF2006}, and the bronze route~\cite{Withanage2021,Rey2021}. Among these, the bronze process—originally developed for multifilamentary wire fabrication~\cite{Godeke_2006,Laurila:APL2010}—has gained renewed attention due to its relatively low synthesis temperature ($\sim$700\,$^\circ$C), which makes it compatible with Cu substrates. In its thin-film adaptation, the electrochemical–thermal synthesis (ETS)  route involves sequential deposition of a Nb diffusion barrier and a Cu--Sn (bronze) precursor layer, followed by annealing to promote solid-state Nb$_3$Sn formation~\cite{Barzi_SUST2015,Lu2022}. Compared to other methods, the ETS route offers several advantages, including low cost, excellent scalability for batch production, and strong adaptability to complex cavity geometries. Furthermore, the process parameters—such as layer composition, thickness, and annealing conditions—can be precisely tuned, enabling controllable and reproducible Nb$_3$Sn film properties.

Despite its potential, applying the bronze route to Cu substrates introduces unique complications. Cu inclusions in thin films can form normal-conducting regions, which degrade RF performance by increasing surface resistance and inducing premature quenching~\cite{Ilyina2019,Withanage2021}. Residual bronze, Sn-rich impurity phases (e.g., Nb$_6$Sn$_5$, NbSn$_2$), and oxygen-related compounds such as Nb oxides (NbO$_x$) and Sn oxides (SnO$_2$) further exacerbate RF losses and limit field stability~\cite{Jaeyel2020155}. Moreover, thermal stress arising from the mismatch in thermal expansion coefficients between Nb$_3$Sn and Cu can induce microcracks within the film or along grain boundaries, but also locally suppress the superconducting transition temperature $T\textsubscript{c}$, thereby further deteriorating the overall superconducting properties~\cite{Godeke_2006}. These challenges demand careful control of plating, annealing, and post-treatment to achieve uniform, phase-pure, and defect-free Nb$_3$Sn coatings on Cu substrates.

In this study, we report the first successful fabrication and RF characterization of a Cu-based Nb$_3$Sn film synthesized via an improved ETS route, specifically tailored for Quadrupole Resonator (QPR) testing. Key process innovations include electropolishing of the Cu substrate, deposition of a dense Nb barrier by magnetron sputtering, optimization of bronze electroplating and high-temperature annealing parameters, and the implementation of a selective chemical etching step to eliminate residual contaminants. The resulting Nb$_3$Sn/Cu sample demonstrated exceptionally low surface resistance at cryogenic temperatures, representing the best performance reported to date for this material system. These advances highlight the feasibility of the ETS route for scalable and cost-effective Nb$_3$Sn/Cu coatings, providing a promising pathway for the next generation of SRF cavity development.

\section{Sample Preparation and Experimental Methods}

\subsection{Overview}

\begin{figure}[htbp]
  \centering
  \includegraphics[width=1.0\textwidth]{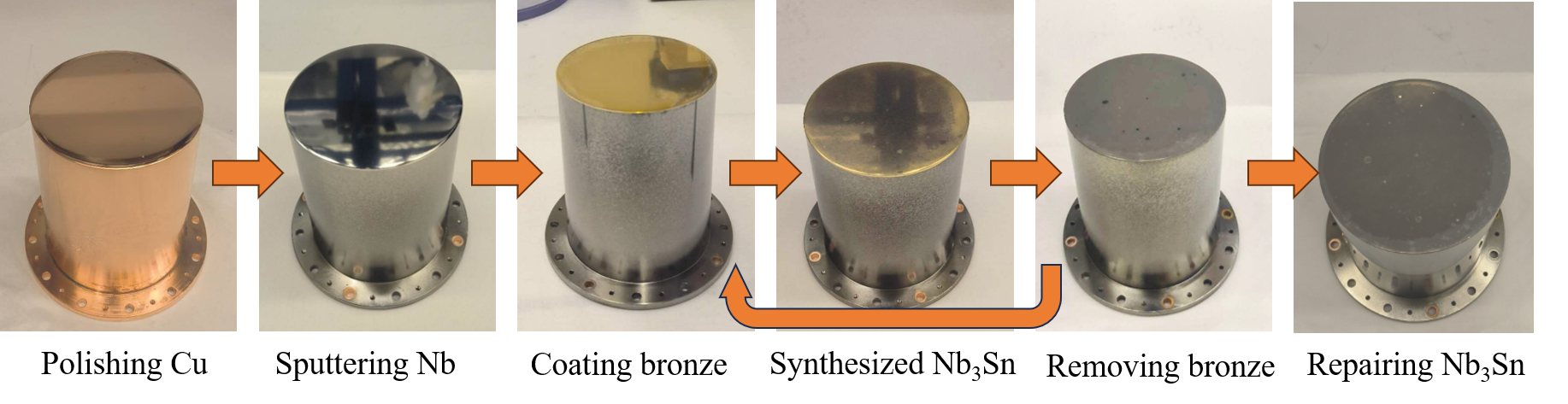}
  \caption{Flowchart of the preparation procedure for the Cu-based Nb$_3$Sn sample.}
  \label{flowchart}
\end{figure}

This section outlines the complete fabrication route and characterization methodology employed for producing Nb$_3$Sn thin films on Cu substrates using the bronze technique. The multi-step procedure—schematically illustrated in Fig.~\ref{flowchart}—includes Cu substrate surface preparation, deposition of the Nb barrier layer, electroplating of the bronze precursor, heat treatment for Nb$_3$Sn phase formation, and final chemical etching to remove top impurity layer. 

Initial attempts revealed that the resulting Nb$_3$Sn films exhibited non-uniform coverage, with some localized uncoated regions (small holes). These defects were suspected to originate from air bubbles trapped on the surface during the immersion of the inverted QPR sample into the bronze plating solution, which inhibited uniform bronze deposition. To address this issue, a repair procedure was implemented. The sample surface was activated using Citranox\textsuperscript{\textregistered} to enhance hydrophilicity, and after the electrolyte was stabilized, it was re-immersed into the electroplating bath at a tilted angle to minimize bubble adhesion. A new bronze layer was deposited, followed by repeated heat treatment and chemical etching. This improved process resulted in a more uniform and continuous Nb$_3$Sn coating over the entire plated region of the QPR sample (with a diameter of about 7.5 cm).

A detailed description of each fabrication step, along with its intended purpose, is summarized in Table~\ref{tab:fabrication}. It should be noted that steps 7-10 were repeated twice for this QPR sample due to repair experiments. The process begins with mechanical and electrochemical polishing to ensure a clean and smooth Cu surface. A dense Nb layer with a thickness of about 10~\textmu m is subsequently sputtered onto the substrate, serving both as a reactive layer for Nb$_3$Sn formation and as a diffusion barrier to prevent interdiffusion between Cu and Nb$_3$Sn. This Nb layer must possess high density and sufficient thickness to prevent the Nb$_3$Sn film from degrading. So, the Nb film was then annealed at 700\,\textcelsius\ for 20\,h to promote grain growth, enhance film density and relieve internal stress. It is then followed by the electroplating of a top bronze precursor layer. Controlled annealing with a slow ramp-up to 700\,\textcelsius\ and holding for 30\,h promotes solid-state diffusion to form the Nb$_3$Sn superconducting phase. Post-treatments, including aqua regia etching and ultrapure water rinsing, remove residual surface bronze impurities, and contaminants. 

\begin{table}[htbp]
\centering
\caption{Fabrication procedure of the Cu-based Nb\textsubscript{3}Sn sample}
\label{tab:fabrication}
\resizebox{\textwidth}{!}{%
\begin{tabular}{@{}cll@{}}
\toprule
Step & Process Description & Purpose \\
\midrule
1 & Mechanical polishing to 100~$\mu$m & Removal of surface damage and reduction of defects  \\
2 & Degreasing with 1\% Citranox\textsuperscript{\textregistered} for 1 hour & Eliminate organic contaminants and grease \\
3 & Electro-polishing (EP) 30~$\mu$m & Achieve a smooth surface with reduced roughness \\
4 & Rinse with ultrapure water & Remove residual acids and particles \\
5 & Sputtering of 10~$\mu$m Nb layer at Siegen & Deposit a superconducting Nb barrier layer \\
6 & High-temperature annealing at 700~°C for 20~h & Enhance film density and relieve internal stress \\
7 & Electroplating of 20~$\mu$m bronze layer & Deposit bronze precursor layer for Nb\textsubscript{3}Sn formation \\
8 & Heat treatment at 700~°C for 30~h  & Form Nb\textsubscript{3}Sn via solid-state diffusion reaction \\
9 & Removal of surface bronze using aqua regia & Expose  Nb\textsubscript{3}Sn layer and remove residual impurities \\
10 & Final rinse with ultrapure water & Ensure clean, residue-free surface for RF testing \\
\bottomrule
\end{tabular}%
}
\end{table}

To evaluate the RF performance of Nb$_3$Sn films, we employ the QPR, a highly sensitive diagnostic tool designed for thin-film characterization under controlled frequency, temperature, and RF field conditions. The QPR measures the surface resistance of the sample using a calorimetric method, achieving an accuracy in the sub-nanoohm range. For details of the method, see Refs.~\cite{Kleindienst_2017,Keckert_2019}.
By analyzing $R\textsubscript{s}$ as a function of temperature, the BCS and residual components of the surface resistance can be extracted via fitting procedures. The measurement uncertainty of $R\textsubscript{s}$ is estimated to be below 5\%~\cite{Keckert_2019}. 

This approach facilitates sensitive and accurate characterization of RF losses in superconducting films under various operational conditions. Operating at 433, 866, and 1300~MHz, the QPR allows direct assessment of frequency-dependent RF losses~\cite{Keckert.S2021,Keckert2021}. Its mechanical stability and precise thermal control down to 1.9~K ensure high measurement reproducibility. By minimizing electric field contributions, the QPR enables measurements dominated by magnetic losses, allowing accurate extraction of key parameters such as $R\textsubscript{s}(B,T)$, the penetration depth $\lambda(T)$, and the maximum RF field $B\textsubscript{max}$. These insights provide essential feedback for establishing the relationship between microstructural features and RF performance, thereby guiding the further refinement of Nb$_3$Sn thin-film fabrication techniques.

\subsection{Substrate Preparation}

The Cu QPR sample used in this study was a monolithic, weld-free structure directly machined from high-purity oxygen-free high-conductivity (OFHC) Cu (top diameter: 75\,mm; height: 100\,mm), with dimensions compatible with standard QPR testing requirements\cite{Keckert.S2021}. To ensure a clean and smooth surface suitable for thin-film deposition, the sample was mechanically polished using SiC sandpapers of progressively finer grit sizes (\#180, \#300, \#500, \#800, \#1200, \#2000, and \#5000) to remove approximately 100\,\textmu m of material.

\begin{figure}[htbp]
    \centering
    \includegraphics[width=1.0\textwidth]{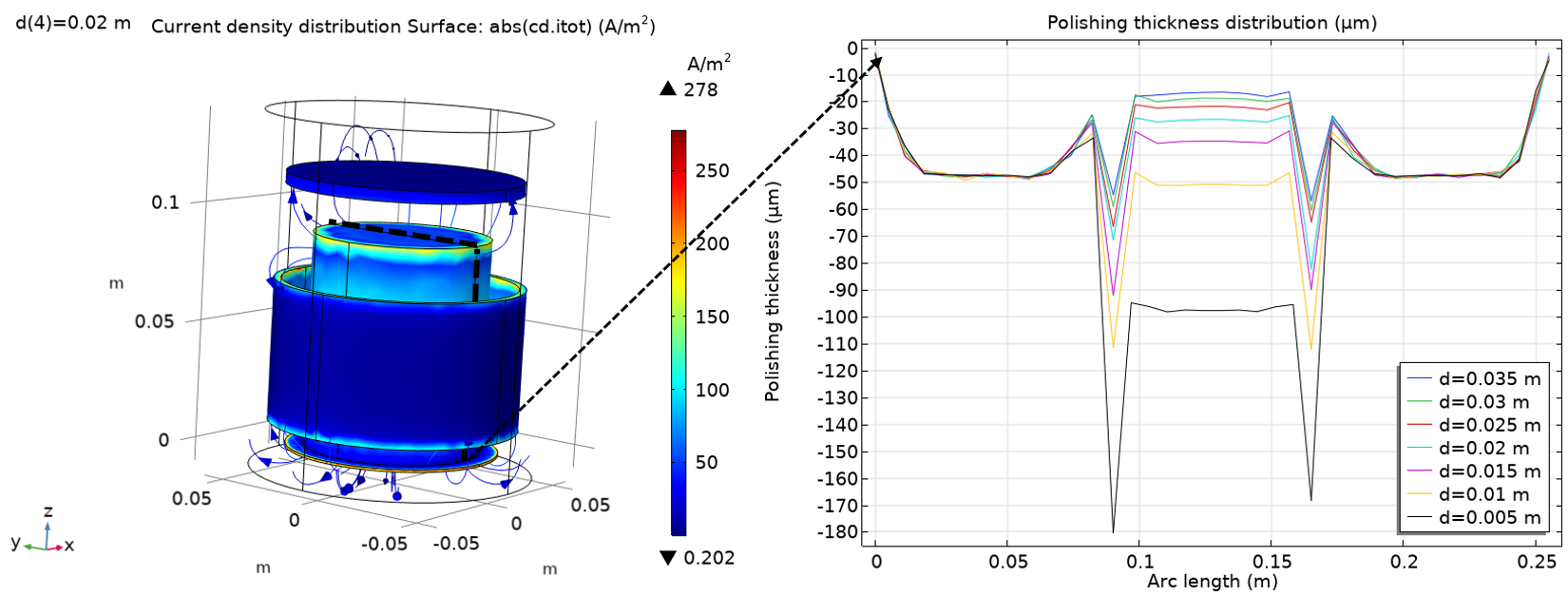}
    \caption{COMSOL Multiphysics simulation of the Cu material removal during electrochemical polishing of the QPR sample, showing improved uniformity with minor edge effects influenced by electrode geometry and electrolyte dynamics.}
    \label{fig:EP_simulation}
\end{figure}

Subsequently, electrochemical polishing (EP) was employed to further improve surface quality, removing an additional $\sim$30\,\textmu m. To optimize the polishing uniformity, a COMSOL Multiphysics simulation was performed to analyze the thickness distribution under various electrode arrangements conditions. The simulation used the ``Secondary Current Distribution'' interface, assuming Tafel kinetics at the anode and cathode surfaces, with exchange current densities of 10\,A/m\textsuperscript{2} and anodic/cathodic transfer coefficients of 0.5. Based on the simulation results (Figure~\ref{fig:EP_simulation}), a custom-designed vertical support fixture was fabricated from chemically resistant polytetrafluoroethylene (PTFE) and polyether ether ketone (PEEK) to achieve uniform material removal. The fixture allowed precise positioning of the Cu sample (as the anode) at the bottom, while disc and strip-shaped Cu cathodes were symmetrically arranged above and around it, with an optimized anode–cathode spacing of 10\,mm.

Electropolishing was carried out at room temperature (20\,$^\circ$C) using an electrolyte composed of phosphoric acid (85\,wt\%) and 1-Butanol ($>99$\,\textrm{wt}\%)
in a 3:2 volume ratio. A constant voltage of 2.1\,V was applied for 2 hours under magnetic stirring at 100\,rpm. After polishing, the sample was thoroughly rinsed with deionized water and dried in a nitrogen atmosphere. The resulting surface exhibited a uniform, mirror-like Cu luster, indicating high polishing quality.

\begin{figure}[htbp]
    \centering
    \includegraphics[width=1.0\textwidth]{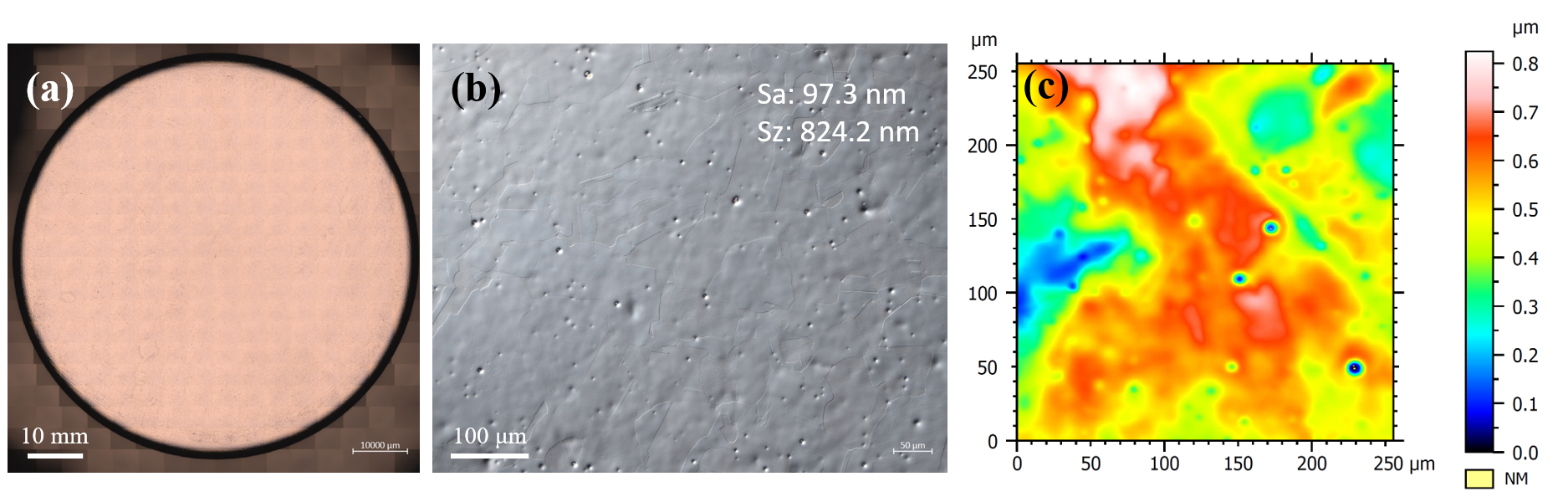}
\caption{Surface images of the Cu substrate after electrochemical polishing: (a) bright-field optical microscope image with a 10 mm scale, (b) dark-field optical microscope image with a 100 \(\mu\)m scale, and (c) LSM topography over a 250 × 250 \(\mu\)m\(^2\) area, showing a smooth surface with an average roughness of ~97.3\,nm and some pitting defects.}
    \label{fig:Cu_Roughness}
\end{figure}

Surface morphology and roughness were characterized using laser scanning microscopy (LSM). As shown in Figure~\ref{fig:Cu_Roughness}, the polished surface exhibits an overall smooth appearance with mean surface roughness (Sa) of approximately 97.3\,nm measured over an area of 250\,$\mu$m\,$\times$\,250\,$\mu$m. However, a significant number of micro-scale surface pits are observed, with typical diameters of several micrometers and depths less than 1\,\textmu m. These defects are suspected to originate from oxygen-bubble formation on the surface during the EP process, which disrupts uniform dissolution. Although the current surface quality meets the basic requirements for thin-film deposition, these surface pits remain a challenge and may adversely affect the coating uniformity. Efforts are ongoing to optimize the EP process parameters and fixture design to mitigate this issue.

\subsection{Nb Reaction Layer Deposition}

The Nb film, serving as both a diffusion barrier and a precursor for subsequent Nb$_3$Sn formation, was deposited on the outer surface of the Cu QPR sample using high-power impulse magnetron sputtering (HiPIMS). The deposition was carried out at a substrate temperature of approximately 180\,°C in an argon atmosphere with a working pressure of 0.02\,mbar and an Ar flow rate of 700\,sccm. A high-purity Nb target (99.99\%) was powered with an average cathode power of 600\,W (equivalent to 6.8\,W/cm$^2$), using a HiPIMS pulsed power supply operating at 1000\,Hz with a pulse width of 100\,$\mu$s. A constant substrate bias of $-50$\,V was applied. The base pressure prior to sputtering was maintained below $7 \times 10^{-7}$\,mbar to ensure a contamination-free environment.

To improve coating uniformity on the curved sidewalls and bottom regions of the sample, the Cu QPR sample was mounted on a custom tiltable holder that oscillated within a ±15° angular range during deposition. This dynamic motion enhanced angular coverage and resulted in a Nb film thickness of approximately 10\,$\mu$m on the top surface. However, due to geometric shadowing and the inherently directional nature of sputtering, the film thickness gradually decreased along the vertical sidewalls, falling below 1\,$\mu$m near the bottom edge. In this region Nb$_3$Sn synthesis does not take place, and one simply must ensure that the cylinder remains superconducting. Since the RF field is very low in this cylinder region and thickness is still far more than the Nb penetration depth. 

\begin{figure}[htbp]
    \centering
    \includegraphics[width=1.0\textwidth]{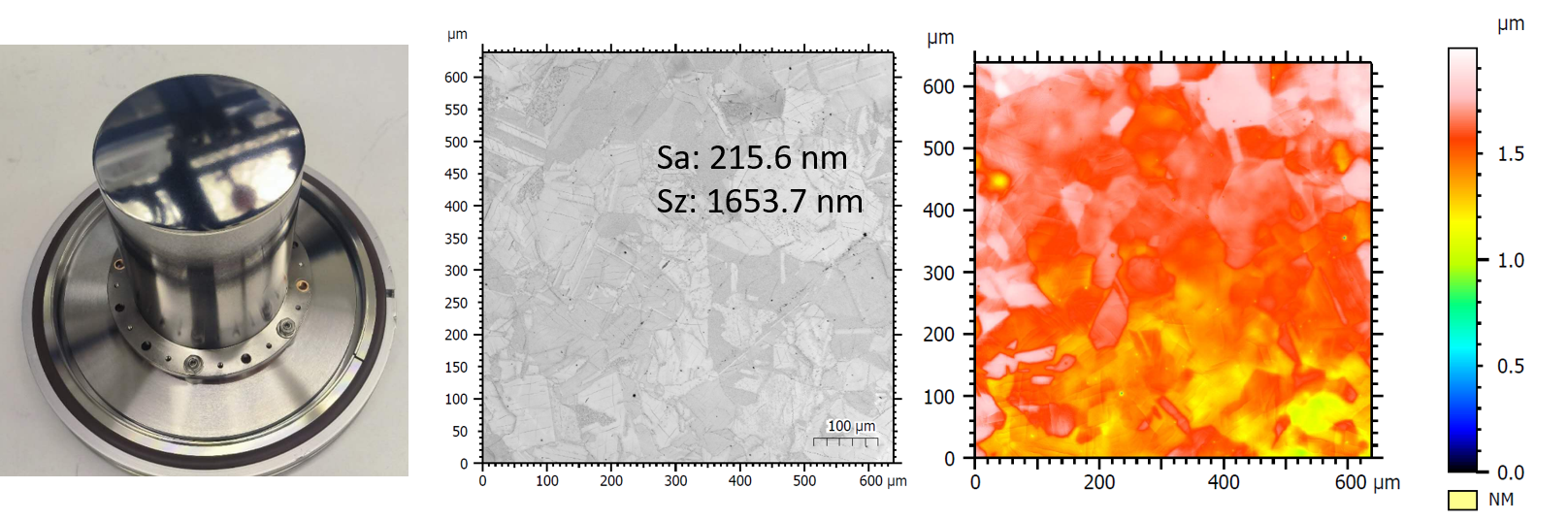}
    \caption{LSM image of the Nb reaction layer surface, showing a smooth, uniform morphology free of cracks, indicative of high-quality film deposition.}
    \label{fig:Nb_roughness}
\end{figure}

Surface morphology and roughness of the deposited Nb film were characterized using LSM. The analysis confirmed a significant reduction in surface pitting compared to the underlying Cu substrate, revealing a continuous and crack-free Nb layer with no visible contamination. As shown in Figure~\ref{fig:Nb_roughness}, the Nb film exhibits a relatively uniform surface morphology with Sa of approximately 215.6\,nm measured over an area of 600\,$\mu$m\,$\times$\,600\,$\mu$m. This level of smoothness is considered suitable for the subsequent Nb$_3$Sn formation.

\subsection{Bronze Electroplating}

To supply the tin source for Nb$_3$Sn formation, a bronze precursor layer was electroplated onto the Nb-coated surface of the QPR sample using an alkaline citrate-based electrolyte. The composition of the plating solution is listed in Table~\ref{tab:bronze_solution}. Since the Nb film on the side and bottom surfaces of the sample is too thin, there is a risk of copper exposure during the bronze-route Nb$_3$Sn coating process, and the presence of exposed copper would lead to excessive RF losses and obscure the intrinsic surface resistance $R\textsubscript{s}$ of the Nb$_3$Sn layer. To avoid this problem, the Nb$_3$Sn coating was applied only to the top surface of the QPR sample, while the existing Nb film on the side and bottom surfaces was left unchanged. This does not impact significantly the calorimetric surface-resistance measurement, since th RF field is much larger on the top surface and decays exponentially along the side walls. 

\begin{table}[htbp]
\centering
\caption{Composition of the bronze electroplating solution~\cite{LU2021129557}}
\label{tab:bronze_solution}
\begin{tabular}{|l|c|c|c|}
\hline
\textbf{Chemical} & \textbf{Formula} & \textbf{Purity} & \textbf{Concentration (g/L)} \\
\hline
Citric acid & C$_6$H$_8$O$_7$ & AR 99.5\% & 180 \\
Potassium stannate & K$_2$SnO$_3$ & AR 95.0\% & 20 \\
Basic copper carbonate & Cu$_2$(OH)$_2$CO$_3$ & AR & 16 \\
Potassium hydroxide & KOH & AR 99.5\% & 130 \\
Monopotassium phosphate & KH$_2$PO$_4$ & AR & 17 \\
\hline
\end{tabular}
\end{table}

\begin{figure}[htbp]
    \centering
    \includegraphics[width=1.0\textwidth]{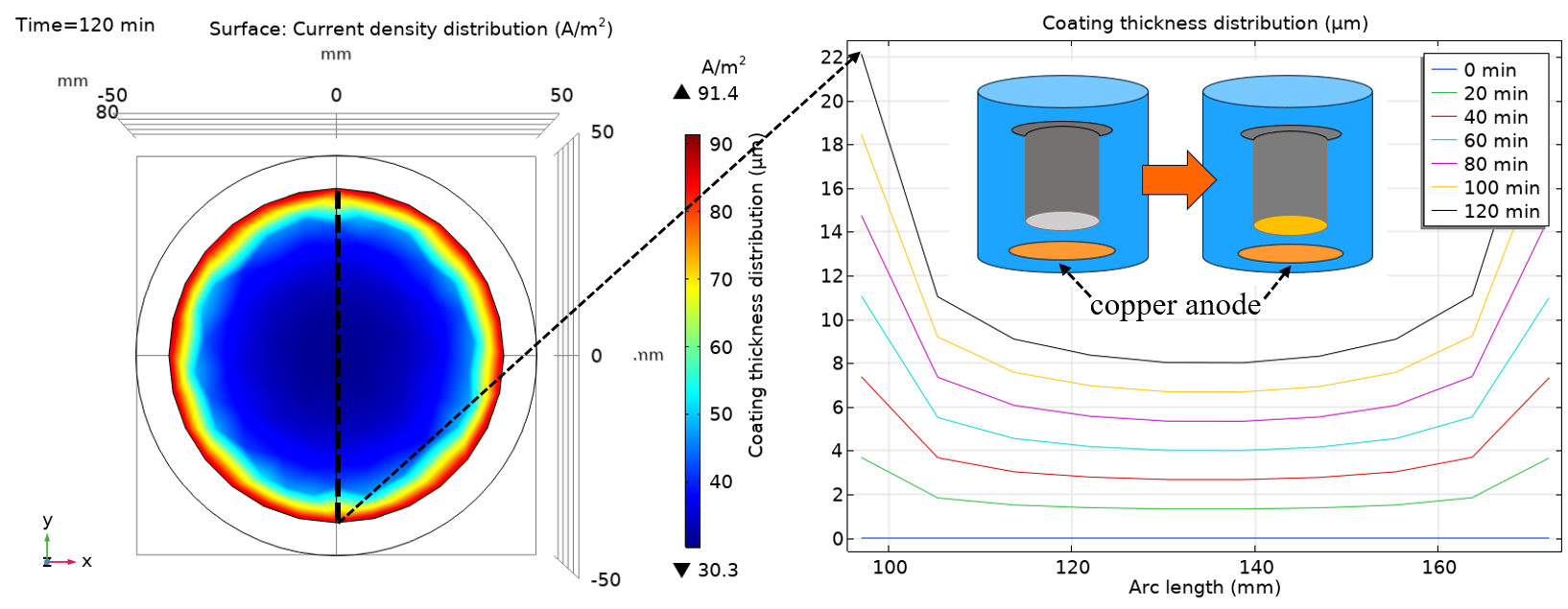}
    \caption{COMSOL simulation of the bronze electroplating thickness distribution on the QPR sample, showing the influence of masking strategy, sample orientation, and electrode geometry on coating uniformity.}
    \label{fig:Bronze_Simulation}
\end{figure}

Similar to the optimization of the EP process, a detailed simulation of electroplating process and the bronze deposition thickness distribution was conducted using COMSOL Multiphysics to optimize the electrode configuration and masking strategy. The simulation results, shown in Figure~\ref{fig:Bronze_Simulation}, the horizontal axis (“arc length”) corresponds to the line-scan distance along the black dashed line indicated on the QPR sample surface. The inset illustrates the relative spatial arrangement of the QPR sample and the copper anode, where the bottom orange circle represents the anode. Besides, the simulation employed the Secondary Current Distribution interface, with Tafel kinetics applied at both the anode (bronze electrode) and the cathode (QPR sample) surfaces. Exchange current densities were set to 10\,A/m\textsuperscript{2}, and both anodic and cathodic transfer coefficients were assumed to be 0.5. The electrolyte was modeled with uniform conductivity, and convection effects were neglected, assuming a charge-transfer-dominated process under steady-state conditions. The primary objective of the simulation was to achieve uniform bronze deposition on the top surface of the QPR sample by optimizing the electrode geometry, spacing, and masking design. To prevent unwanted deposition, the sidewalls and bottom surface---where the niobium barrier layer (below 1~$\mu$m) was insufficiently thick---were isolated from the electrolyte during the plating process. 

Based on the simulation insights, the QPR sample was mounted upside down with its top surface partially immersed in the electrolyte. The sidewalls and bottom were masked with chemically resistant tape to block deposition in these areas. A horizontal Cu anode disk was positioned directly below the sample 20~mm to provide uniform current distribution and minimize anode sludge contamination. Electroplating was performed in galvanostatic mode at a constant current density of 5\,mA/cm$^2$ for 2 hours, at a bath temperature of 20\,°C. This process yielded a bronze layer of approximately 10\,\textmu m on the top surface. 

\begin{figure}[htbp]
    \centering
    \includegraphics[width=1.0\textwidth]{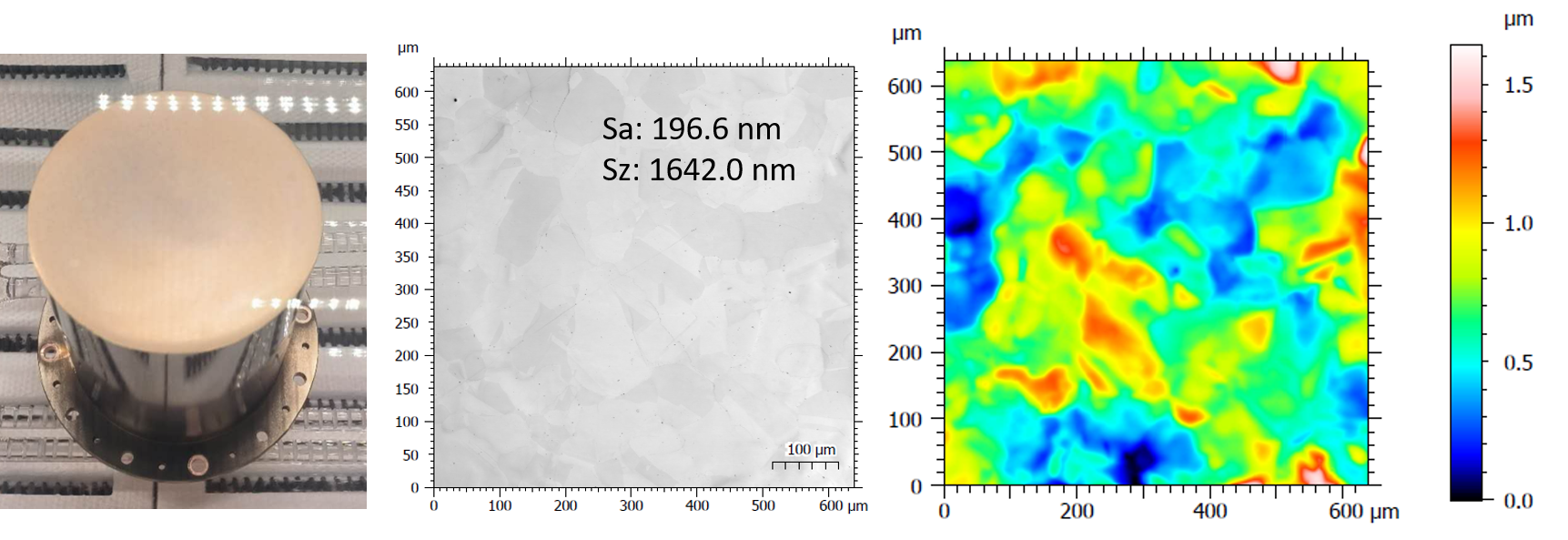}
    \caption{LSM image of the bronze-coated surface on the Nb-coated QPR sample. The surface shows a smooth, uniform morphology with no visible cracks or contamination.}
    \label{fig:Bronze_Roughness}
\end{figure}

After plating, the sample was rinsed with ultrapure water and dried under a nitrogen stream. The surface morphology and roughness of the bronze layer were examined using LSM. As shown in Figure~\ref{fig:Bronze_Roughness}, the bronze coating exhibited a uniform and crack-free morphology with no apparent particulate contamination, indicating the effectiveness of the optimized electroplating setup. The measured arithmetic Sa was 196.6\,nm measured over an area of 600\,$\mu$m\,$\times$\,600\,$\mu$m, which is acceptable for the subsequent Nb$_3$Sn diffusion reaction. Interestingly, after bronze electroplating, the Sa decreased while the Sz remained essentially unchanged compared with the previous Nb sputtering step. This can be explained by the deposition of bronze into micro-scale depressions, which reduces the overall average roughness, whereas the extreme surface peaks and valleys remain largely unaffected.

\subsection{Heat Treatment for Nb$_3$Sn Synthesis}

\begin{figure}[htbp]
    \centering
    \includegraphics[width=1.0\textwidth]{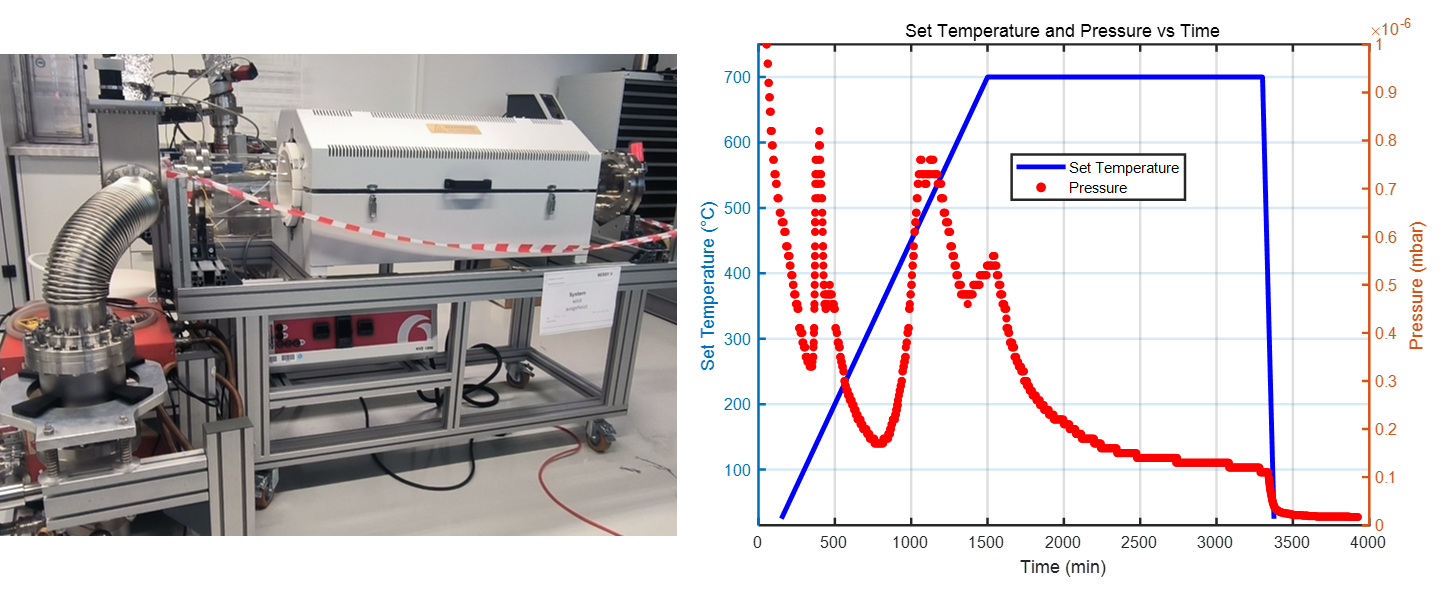}
    \caption{Photograph of the horizontal three-zone KVZ~12/1200 furnace with the pumping system, and a plot of the programmed set temperature and measured chamber pressure versus time during Nb$_3$Sn heat treatment.}
    \label{fig:T-pressure}
\end{figure}

All heat treatments were conducted in a horizontal three-zone KVZ~12/1200 tube furnace with a maximum operating temperature of 1200~\(^\circ\)C and a total heated length of 1200~mm. A 150~mm diameter quartz tube served as the reaction chamber, and the samples were placed at the center of the uniform-temperature zone. The furnace was evacuated using a combination of a dry scroll pump, a turbomolecular pump, and a cryopump. A base pressure below \( 5 \times 10^{-8} \,\mathrm{mbar} \) was achieved at room temperature, and the pressure at \( 700\,^\circ\mathrm{C} \) was maintained below \( 5 \times 10^{-7} \,\mathrm{mbar} \). Figure~\ref{fig:T-pressure} shows the actual furnace and pumping system, as well as the programmed temperature profile and the measured pressure variation during a representative heat treatment process.

Multilayer bronze/Nb/Cu QPR sample was annealed under high vacuum to induce the formation of Nb$_3$Sn via interdiffusion of Sn and Nb, while minimizing undesired Sn penetration through the Nb diffusion barrier. The heat treatment involved a slow ramping process, in which the QPR sample was heated to 700 °C at a rate of 0.5 °C/min and held for 30 hours. Owing to the use of a pre-deposited bronze layer as the precursor, the conventional low-temperature alloying step ($\approx 210~^\circ$C) and intermediate Sn diffusion step ($\approx 400~^\circ$C) were omitted~\cite{Lu_2025}. This direct high-temperature growth significantly reduced the total annealing time and overall processing cost, while enabling the formation of the Nb$_3$Sn phase through Sn diffusion and reaction at the Nb interface. After annealing, the QPR sample was cooled to room temperature by natural cooling.

\subsection{Nb$_3$Sn surface etching}

Post-annealing surface etching is crucial for removing residual bronze, surface contaminants, and reaction byproducts, thereby exposing a clean and pure Nb$_3$Sn layer suitable for subsequent RF characterization. A series of chemical formulations—including phosphoric acid, nitric acid, and hydrogen peroxide—were systematically evaluated for surface cleaning, among which freshly prepared aqua regia (a 3:1 volumetric mixture of 37\,wt\% HCl and 68\,wt\% HNO$_3$) demonstrated superior performance. This optimized formulation effectively removes surface impurities without attacking or degrading the Nb$_3$Sn film.
 
\begin{figure}[htbp]
    \centering
    \includegraphics[width=1.0\textwidth]{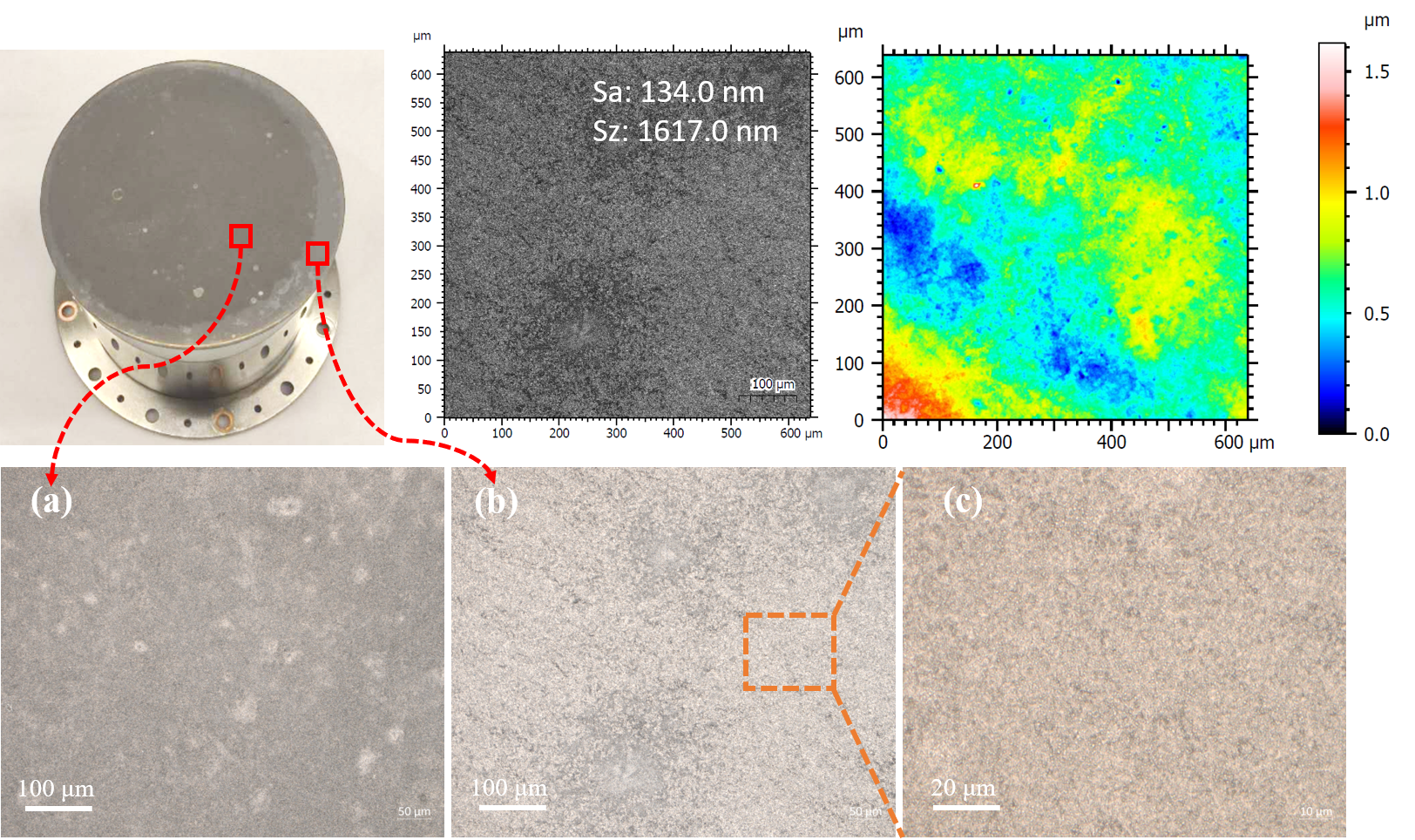}
  \caption{
LSM image of the Nb$_3$Sn surface after surface etching, showing the surface morphology of different regions. 
(a) Microstructure of the high magnetic field region beneath the QPR poles, exhibiting darker contrast. 
(b) Microstructure of the lighter edge region. 
(c) Magnified view of a small portion of the edge region shown in (b).
}
    \label{fig:Nb3Sn_Roughness}
\end{figure}

After chemical etching at 25\,$^\circ$C for 20 minutes, the samples were examined using LSM to characterize the surface roughness and morphology of the Nb$_3$Sn layer. As shown in Figure~\ref{fig:Nb3Sn_Roughness}, the surface exhibits noticeable voids and irregular features caused by pits on the underlying Cu substrate introduced during the electrochemical polishing process. The arithmetic Sa was measured to be 134.0\,nm over an area of 600\,$\mu$m\,$\times$\,600\,$\mu$m. A comparison between the same area before and after etching could further clarify the correlation between surface voids and the underlying pits introduced during Cu electrochemical polishing. Such surface defects, especially the pits, may locally degrade RF performance. 

Figures~\ref{fig:Nb3Sn_Roughness}(a--c) show optical images revealing compositional inhomogeneity. The darker central region corresponds to near-stoichiometric Nb$_3$Sn, while the brighter corners indicate higher Sn content. This assignment is based on the estimated local current density during bronze electroplating: the edges experience higher current density, leading to a higher Sn concentration in the bronze precursor. This is attributed to higher local current density during bronze electroplating at the sample edges, resulting in increased Sn content in those regions. It should be noted that the QPR primarily probes the surface resistance in the crescent-shaped region directly beneath the pole shoes of the resonator. While the edges of the sample are richer in Sn, LSM images also reveal a few tin-rich spots within this crescent-shaped region under the poles. Therefore, the compositional inhomogeneity cannot be completely neglected and may contribute to variations in the measured RF surface resistance.

Following the LSM inspection, the samples were promptly transferred to a class-100 cleanroom (ISO 5) for thorough rinsing with ultrapure deionized (DI) water to eliminate any residual contaminants. The rinsed samples were then dried under a flow of high-purity nitrogen gas. After drying, the samples were carefully mounted onto customized flanges compatible with the QPR system, ensuring minimal handling contamination. Finally, the assembled samples were transferred under clean conditions to the QPR test setup for RF measurements.


\section{Results and Discussion}

To evaluate the superconducting and radio-frequency (RF) performance of the Nb$_3$Sn/Cu film, a comparative study was performed using both the baseline Nb/Cu sample and the ETS Nb$_3$Sn/Cu sample. All RF characterizations were carried out with the Quadrupole Resonator (QPR). In parallel, microstructural and compositional analyses were conducted on flat witness samples (30\,mm $\times$ 20\,mm $\times$ 3\,mm) fabricated under identical processing conditions as the QPR samples. 

\subsection{RF Surface Resistance of Baseline Nb/Cu QPR Sample}

\begin{figure}[htbp]
\centering
\includegraphics[width=1.0\textwidth]{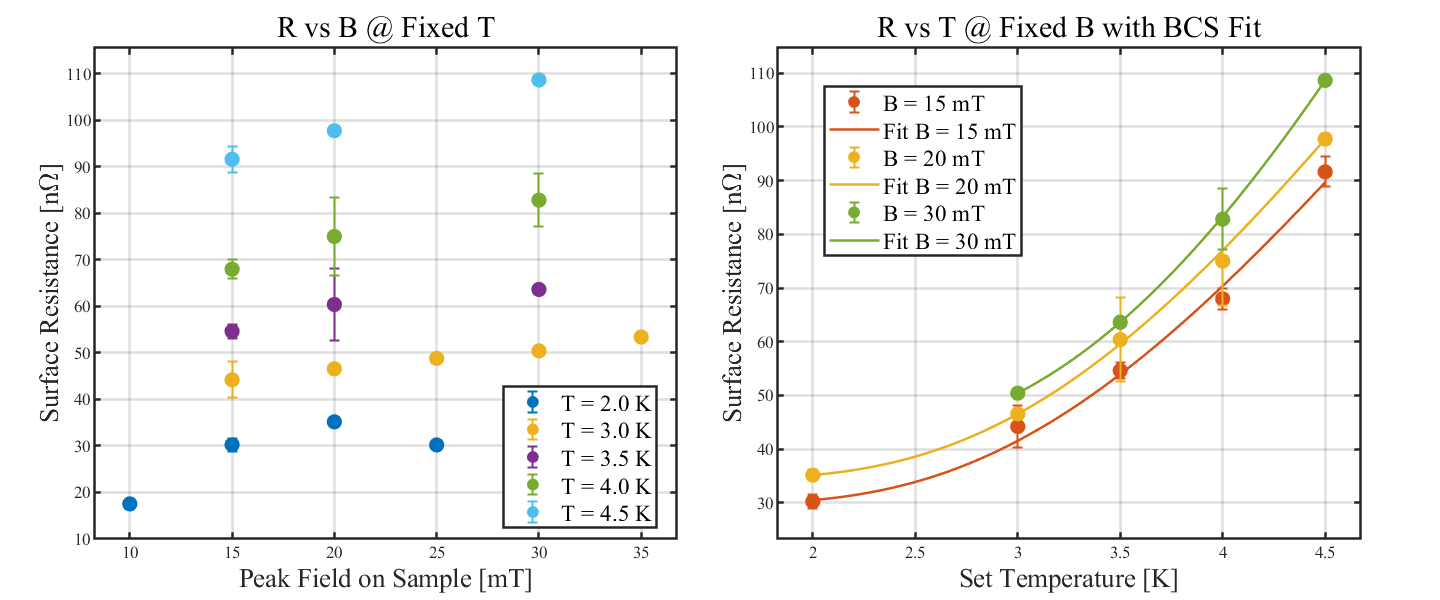}
\caption{Surface resistance $R\textsubscript{s}$ of the baseline Nb/Cu sample measured at 412~MHz: (Left) $R\textsubscript{s}$ vs. peak RF magnetic field; (Right) $R\textsubscript{s}$ vs. temperature.}
\label{fig:baseline_nb_rs_413}
\end{figure}

\begin{figure}[htbp]
\centering
\includegraphics[width=1.0\textwidth]{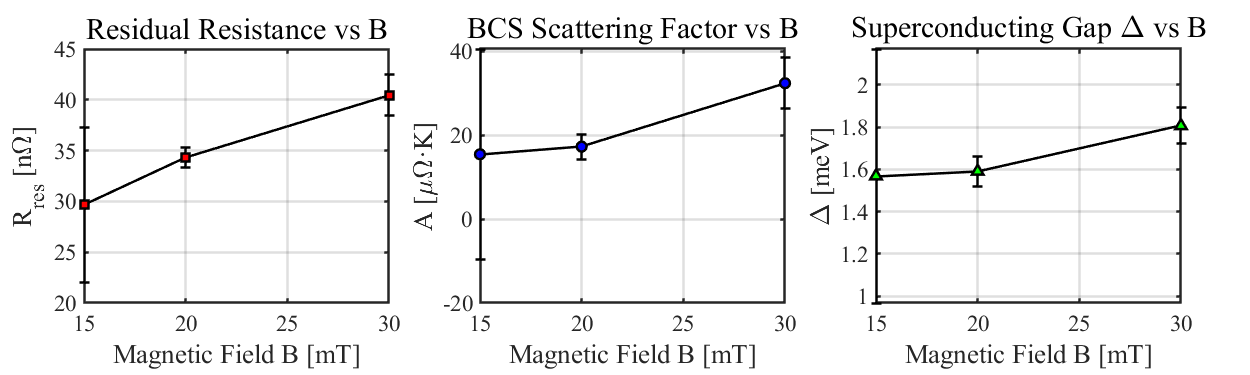}
\caption{Extracted BCS parameters from Nb/Cu sample $R\textsubscript{s}$(T) curves at 412 MHz as a function of RF fields level.}
\label{fig:Nb-Fitting_412}
\end{figure}

The reference Nb/Cu sample was tested to establish a baseline for RF performance evaluation. The RF surface resistance $R\textsubscript{s}$ was measured at 412~MHz. As shown in Figure~\ref{fig:baseline_nb_rs_413}, the sample exhibits expected temperature and field dependencies, with a surface resistance of approximately 90\,n$\Omega$ at 4.5\,K and 15\,mT. This level of performance is consistent with typical sputtered Nb thin films on Cu substrate, reflecting medium-grade RF quality. To further evaluate the superconducting properties of the reference Nb/Cu sample, the measured $R\textsubscript{s}(T)$ curves under different RF field levels were fitted to extract key parameters. The RF surface resistance $R\textsubscript{s}(T)$ was fitted using a simplified BCS-based model, expressed as:
\begin{equation}
R_s(T) = R_{\mathrm{res}} + A \cdot \frac{\omega^2}{T} \exp\left(-\frac{\Delta}{k_B T}\right),
\end{equation}
where $R_{\mathrm{res}}$ is the residual resistance, $A$ is a fitting parameter related to BCS scattering, $\Delta$ is the superconducting energy gap, $\omega$ is the angular frequency of the RF field, and $k_B$ is the Boltzmann constant. The fitting of the measured $R\textsubscript{s}(T)$ curves under different RF field levels allowed extraction of $R_{\mathrm{res}}$, $A$, and $\Delta$. As shown in Figure~\ref{fig:Nb-Fitting_412}, these parameters provide important insights into the material's RF performance and serve as a baseline for comparison with the Nb$_3$Sn-coated sample.

\begin{figure}[htbp]
\centering
\includegraphics[width=1.0\textwidth]{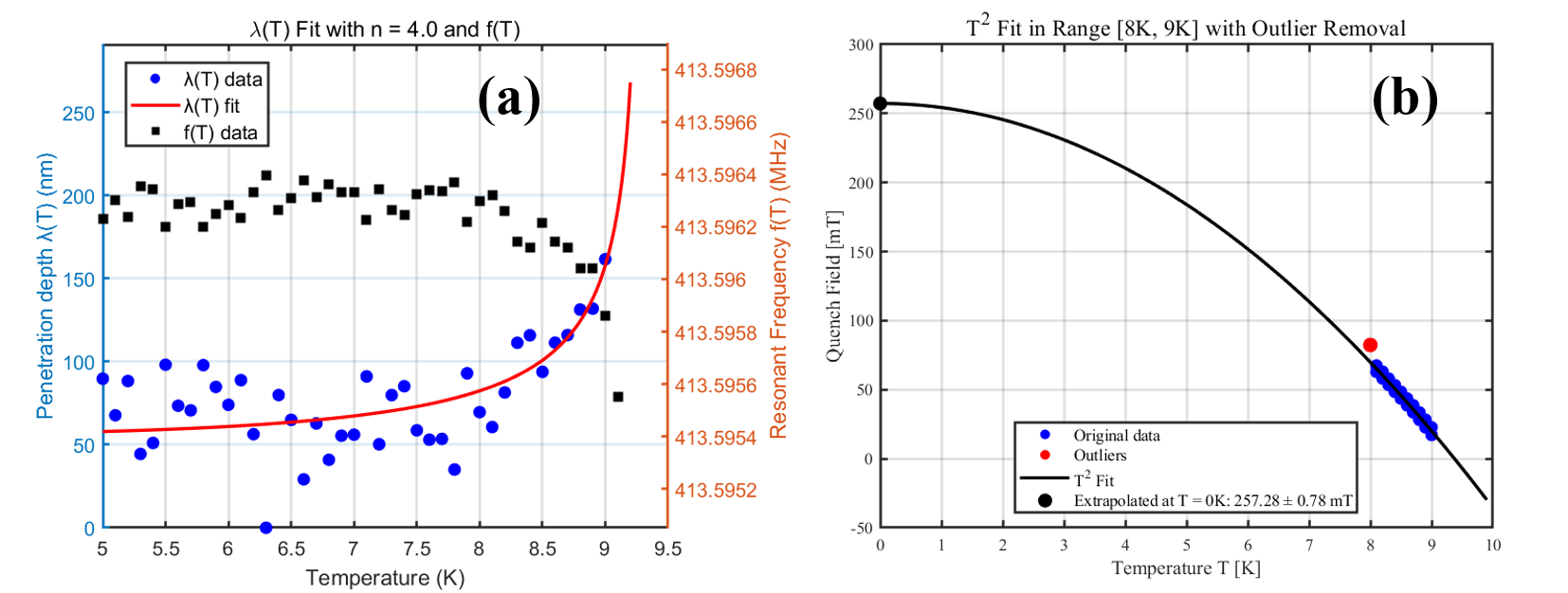}
\caption{Superconducting transition and RF quench characteristics of the baseline Nb/Cu sample. 
(a) Temperature dependence of $\lambda(T)$ and $f(T)$ with fit to $\lambda(T)$. 
(b) Extrapolated quench field at 0\,K using a $T^2$ fit, yielding $\sim$257\,mT.}
\label{fig:nb_tc_quench}
\end{figure}

The superconducting transition temperature of the Nb/Cu sample is extracted from the temperature dependent RF penetration depth $\lambda(T)$, shown in Figure~\ref{fig:nb_tc_quench}a. It exhibits a sharp transition centered at approximately 9.2\,K, consistent with that of pure niobium. The extrapolated RF quench field, obtained from a $T^2$ fit to the temperature-dependent quench field data above 8\,K (Figure~\ref{fig:nb_tc_quench}b), is estimated to be approximately 257\,mT at 0\,K. This value is close to the theoretical superheating field of niobium ($\sim$240\,mT), indicating that the film approaches the intrinsic performance limit of bulk Nb. These results demonstrate that the sputtered Nb/Cu sample exhibits stable and reproducible RF performance consistent with standard Nb/Cu technology. It therefore provides a reliable benchmark for evaluating the performance of the ETS Nb$_3$Sn/Cu sample.

\subsection{RF Surface Resistance of the Cu-Based Nb$_3$Sn QPR Sample}

The Cu-based Nb$_3$Sn film fabricated via the ETS was characterized at 413 and 845 MHz to assess its superconducting and RF properties. Compared to the Nb/Cu baseline, the Nb$_3$Sn sample exhibited a lower surface resistance at temperatures above 4\,K. However, at 2\,K, its surface resistance was slightly higher, reflecting potential material or interface limitations at low temperature. Notably, similar behavior has been observed for Nb$_3$Sn cavities fabricated by vapor diffusion, where elevated low-temperature Rs and field-dependent losses have been linked to non-ideal Sn supply, surface defects (e.g. Sn droplets), and compositional inhomogeneity~\cite{Chen2024, Pudasaini2021}. These results suggest that while the bronze route Nb$_3$Sn/Cu coating shows promise, further optimization of the fabrication process is necessary to achieve consistently superior performance across the full operating temperature range.

\begin{figure}[htbp]
\centering
\includegraphics[width=1.0\textwidth]{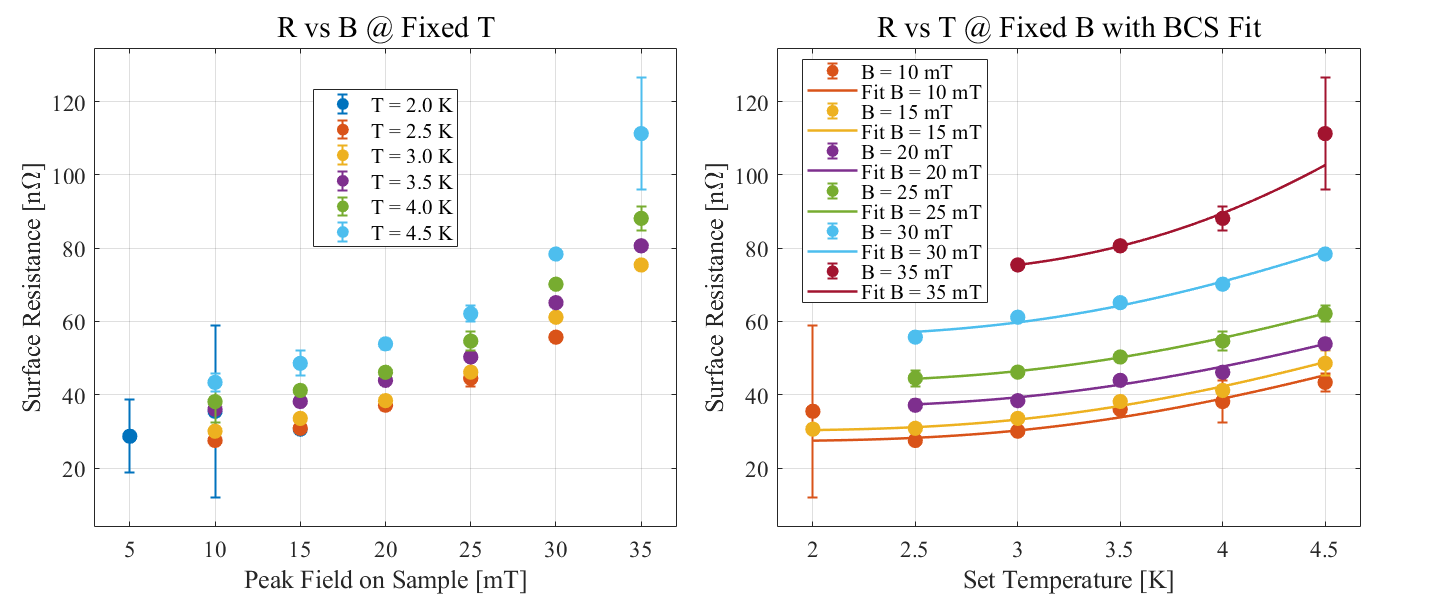}
 \caption{Surface resistance $R\textsubscript{s}$ of the Cu-based Nb$_3$Sn sample measured at 413~MHz. (Left) $R\textsubscript{s}$ vs. peak RF magnetic field. (Right) $R\textsubscript{s}$ vs. temperature.}
\label{fig:nb3sn_rs_413}
\end{figure}

\begin{figure}[htbp]
\centering
\includegraphics[width=1.0\textwidth]{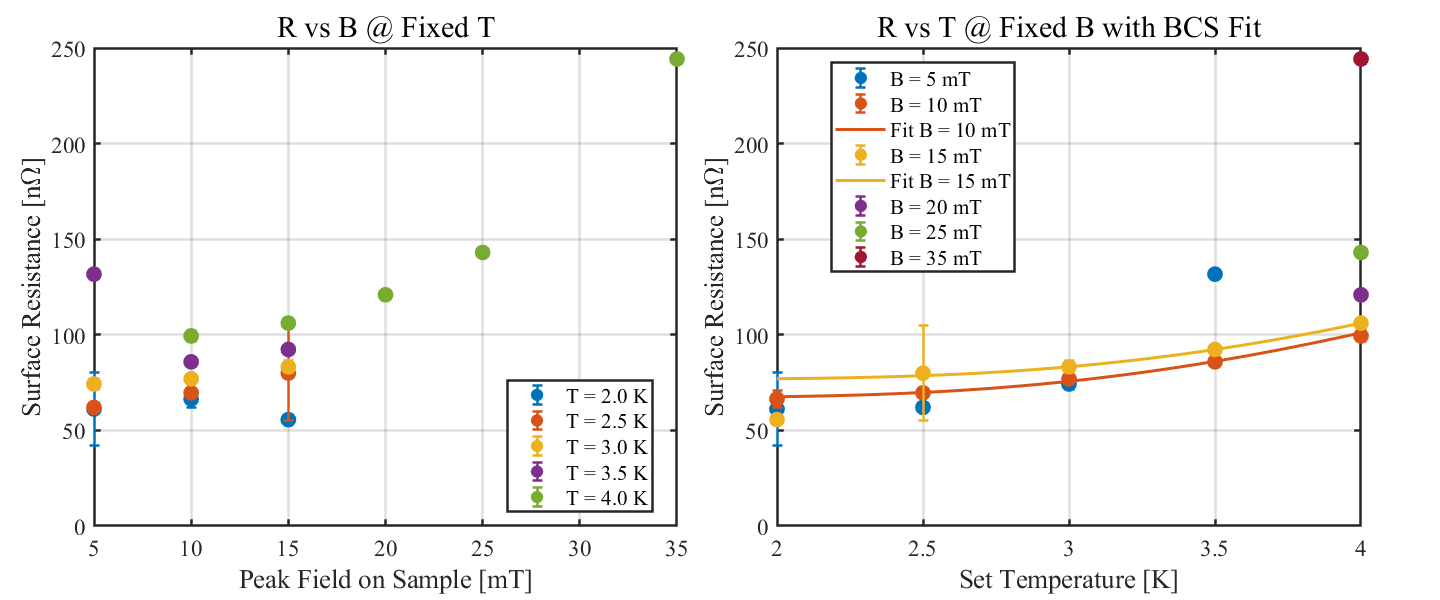}
\caption{Surface resistance of the Cu-based Nb$_3$Sn sample at 845~MHz. (Left) $R\textsubscript{s}$ vs. peak RF magnetic field. (Right) $R\textsubscript{s}$ vs. temperature.}
\label{fig:nb3sn_rs_845}
\end{figure}

\begin{figure}[htbp]
\centering
\includegraphics[width=1.0\textwidth]{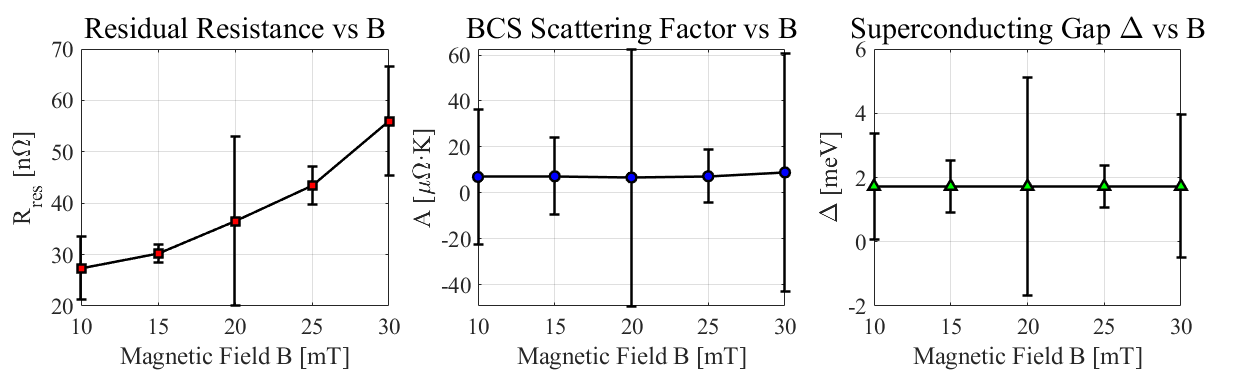}
\caption{Extracted BCS Parameters from Cu-based Nb$_3$Sn sample $R\textsubscript{s}$(T) Curves at 413 MHz under varying RF fields.}
\label{fig:Nb3Sn-Fitting_413}
\end{figure}

Figures~\ref{fig:nb3sn_rs_413} and~\ref{fig:nb3sn_rs_845} present the surface resistance $R\textsubscript{s}$ of the Cu-based Nb$_3$Sn sample measured at 413\,MHz and 845\,MHz, respectively. At 4.5\,K and 15\,mT, the minimum $R\textsubscript{s}$ at 413\,MHz reached 43.4\,n$\Omega$, significantly lower than the sputtered Nb/Cu reference (90\,n$\Omega$) and comparable to early QPR results for Nb$_3$Sn coatings fabricated on Nb substrates via tin vapor diffusion method~\cite{calatroni2023beyond}. At 845\,MHz and 4.0\,K, the $R\textsubscript{s}$ increased to 105\,n$\Omega$ at 15\,mT. Across both frequencies, $R\textsubscript{s}$ exhibits a strong and continuous field-dependent increase throughout the measured RF field range, indicating pronounced $Q$-slope behavior, with a slope exceeding 2 times that of the Nb/Cu sample over the measured B-field range. This degradation is attributed to scattering centers and microstructural imperfections, as evidenced by XRD and EDS analyses (Figure~\ref{fig:Nb3Sn_characterization}, \ref{fig:Nb3Sn_EDX}), which reveal impurity phases such as Nb$_6$Sn$_5$ or NbSn$_2$ arising from localized Sn-rich regions in the Nb$_3$Sn film, as well as occasional Cu inclusions and surface contamination introduced during the coating process.
These imperfections collectively deteriorate the coating quality and are responsible for the pronounced $Q$-slope observed in the RF performance.

To characterize the superconducting behavior of the Cu-based Nb$_3$Sn sample, the $R\textsubscript{s}(T)$ curves measured at 413 MHz under various RF magnetic fields were fitted using the same BCS-based model. As shown in Figure~\ref{fig:Nb3Sn-Fitting_413}, the fitting yields the residual resistance $R{\textsubscript{res}}$, the BCS scattering parameter $A$, and the superconducting energy gap $\Delta$ for each field level. It should be noted that the $R\textsubscript{s}(T)$ fit for Nb$_3$Sn suffers from significant data instability and model limitations, resulting in large errors. Compared to the Nb/Cu reference sample, the Nb$_3$Sn coating exhibits a lower $R{\textsubscript{res}}$ and a larger energy gap $\Delta$, indicating superior RF performance and stronger superconducting pairing. 

\begin{figure}[htbp]
\centering
\includegraphics[width=1.0\textwidth]{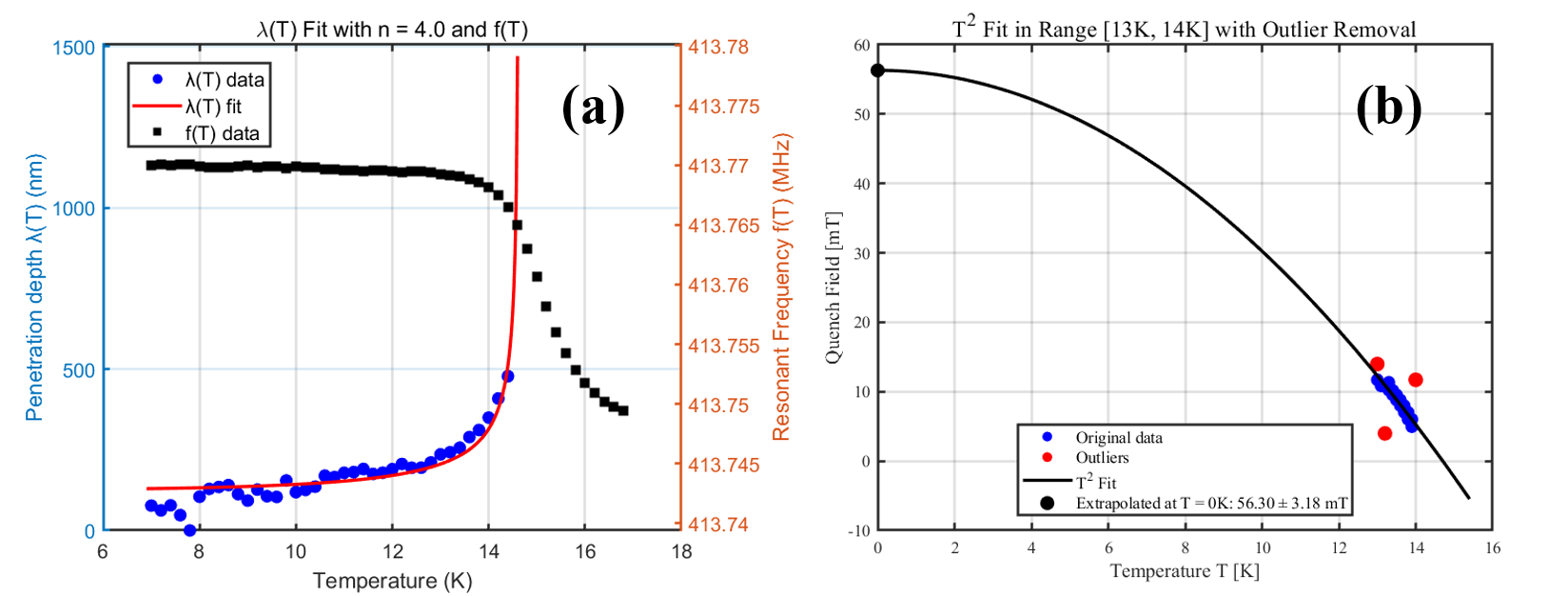}
\caption{Superconducting transition and quench field characteristics of the Cu-based Nb$_3$Sn QPR sample. 
(a) Temperature dependence of $\lambda(T)$ and $f(T)$ with fit to $\lambda(T)$. 
(b) Extrapolated quench field at 0\,K using a $T^2$ fit, yielding $\sim$56.3\,mT.}
\label{fig:nb3sn_tc_quench}
\end{figure}

The RF-detected superconducting transition temperature $T\textsubscript{c}$ of the Cu-based Nb$_3$Sn sample, shown in Figure~\ref{fig:nb3sn_tc_quench}a, was approximately 14.6\,K. Compared to the sharp transition of the Nb/Cu baseline sample (Figure~\ref{fig:nb_tc_quench}a), the transition in the Nb$_3$Sn film appears slightly less sharp. 
This may be partly due to the fitting procedure, but is also consistent with the intrinsic behavior of Nb$_3$Sn, whose $T\textsubscript{c}$ varies with local Sn content, such that regions with slightly lower Sn exhibit reduced $T\textsubscript{c}$. Additional effects such as stress in the film can further reduce $T\textsubscript{c}$ to roughly 80--90\% of its ideal value, contributing to the observed transition width. The quench behavior, presented in Figure~\ref{fig:nb3sn_tc_quench}b, was analyzed using a $T^2$ fit to the temperature-dependent quench field data above 13\,K, yielding an extrapolated intrinsic quench field of 56.3\,mT at 0\,K. Although this value is significantly lower than the theoretical superheating field of Nb$_3$Sn, which exceeds 400\,mT, it is still above the lower critical field \( H_{c1} \approx 38\,\mathrm{mT} \), indicating that the quench is not due to vortex penetration. The reduced quench field likely reflects extrinsic limitations imposed by compositional inhomogeneity, Sn-rich regions, or structural defects such as weakly connected grain boundaries and localized non-superconducting phases. These features can give rise to premature thermal runaway or vortex entry under RF excitation, thereby limiting the achievable surface field well below the intrinsic material limit.

\begin{table}[htbp]
\centering
\caption{Material parameters derived from surface resistance $R\textsubscript{s}(T)$, penetration depth, and quench field for Cu/Nb and Cu/Nb$_3$Sn QPR samples measured at 412--413\,MHz and 15--20\,mT, together with theoretical reference values for bulk Nb and stoichiometric Nb$_3$Sn.}
\renewcommand{\arraystretch}{1.2}
\resizebox{\textwidth}{!}{%
\begin{tabular}{ccccc}
\hline
\textbf{Symbol} & \textbf{Nb/Cu} & \textbf{Nb$_3$Sn/Cu} & \textbf{Nb (theory)} & \textbf{Nb$_3$Sn (theory)} \\
\hline
$\lambda(0)$ [nm]        & $53.89 \pm 6.61$ & $126.09 \pm 10.77$ & $\sim 40$  & $80$--$135$ \\
$\ell$ [nm]              & $20.7 \pm 8.7$   & $2.5 \pm 0.6$      & $>100$ (clean) & $5$--$20$ \\
RRR                      & $6.5 \pm 2.7$    & $0.8 \pm 0.2$      & $200$--$300$ & $2$--$5$ \\
$T\textsubscript{c}$ [K] & $9.30 \pm 0.11$  & $14.63 \pm 0.09$   & $9.25$ & $18.0$ \\
$R\textsubscript{res}$ [$\mathrm{n}\Omega$] & $34.34 \pm 0.97$ & $30.20 \pm 1.77$ & $<10$  & $<10$  \\
$A_{\mathrm{BCS}}$ [$\mu\Omega\cdot\mathrm{K}$] & $17.24 \pm 2.96$ & $7.18 \pm 16.86$ & $10$--$20$ & $5$--$10$ \\
$\Delta(0)/k\textsubscript{B} T\textsubscript{c}$ & $1.59 \pm 0.07$ & $1.72 \pm 0.82$ & $1.85$  & $2.2$  \\
$B\textsubscript{quench}(0)$ [mT] & $257.28 \pm 0.78$ & $56.30 \pm 3.18$ & $200$--$240$ & $400$--$440$ \\
\hline
\end{tabular}%
}
\label{tab:material_parameters}
\vspace{0.5em}
\footnotesize \textit{Note.} Theoretical values are typical literature ranges for bulk Nb and stoichiometric Nb$_3$Sn~\cite{Godeke2006_LBNL62140}. 
\end{table}

Table~\ref{tab:material_parameters} summarizes the superconducting parameters extracted from $R\textsubscript{s}(T)$ fitting, penetration depth analysis, and quench field measurements for the Nb/Cu and Nb$_3$Sn/Cu QPR samples. For the Nb$_3$Sn sample, the measured penetration depth is larger, the mean free path ($\ell$) is shorter, and the RRR is lower than the ideal theory values, indicating stronger electronic scattering and higher disorder, likely introduced by the Cu substrate and bronze-route deposition. 
Although the critical temperature is close to that of stoichiometric Nb$_3$Sn, the large uncertainty in the extracted energy gap ratio suggests spatial variations in superconducting properties. The reduced quench field, relative to the theoretical superheating field, likely reflects local material inhomogeneities, where regions that transition to the normal-conducting state cannot be stabilized. 
This highlights the challenges in achieving high-quality Nb$_3$Sn coatings on copper and the need for improved control over composition and microstructure for SRF applications.

\begin{table}[htbp]
\centering
\caption{Representative $R\textsubscript{s}$ values and quench fields for Nb\textsubscript{3}Sn coatings at 4.5\,K and $\sim$15\,mT, extracted from figures in this work and the cited references.}
\label{tab:comparison}
\begin{tabular}{lccc}
\toprule
\textbf{Method} & \textbf{Substrate} & \textbf{$R\textsubscript{s}$ (n$\Omega$)} & \textbf{$B_\mathrm{quench}$ (mT)} \\
\midrule
Bronze route (this work) & Cu & \textbf{43.4} & 56.3 \\ 
HiPIMS + annealing~\cite{calatroni2023beyond} & Cu & 50--60 & -- \\
DCMS~\cite{calatroni2023beyond} & Cu & $>$100 & -- \\
Sn-vapor diffusion (2017)~\cite{Keckert:SRF2017} & Nb & $\sim$50 & 200$\pm$ 5 \\
Sputtering (2024)~\cite{Fonnesu2024} & Nb & $\sim$20 & $>$70 \\
\bottomrule
\end{tabular}
\end{table}

To contextualize the RF performance of the present Cu-based Nb\textsubscript{3}Sn coatings, Table~\ref{tab:comparison} summarizes representative $R\textsubscript{s}$ values at 4.5\,K and $\sim$15\,mT obtained from different Nb\textsubscript{3}Sn coating methods on both Cu and Nb substrates. These values were extracted from figures and data reported in recent literature~\cite{calatroni2023beyond,Keckert:SRF2017}. The table provides a comparative view of performance across diverse deposition techniques, including HiPIMS, DCMS, Sn-vapor diffusion, and optimized sputtering, highlighting both typical $R\textsubscript{s}$ and quench behavior.
Among the tested techniques, the bronze route currently exhibits the lowest surface resistance on Cu substrates, demonstrating its potential as a promising pathway for SRF applications operating at 4.5~K. Although the measured superconducting transition temperature ($T_\mathrm{c} \approx 14.6$\,K) is lower than the ideal stoichiometric Nb\textsubscript{3}Sn value of 18~K, this reduction is consistent with bronze-route films, where slight Sn inhomogeneity and lattice mismatch-induced stress can locally suppress $T_\mathrm{c}$\cite{Fonnesu2023,ZhuLiang2022}.
In addition, residual impurity phases and oxygen-rich regions—likely originating from insufficient diffusion barriers or suboptimal annealing—can degrade both $T\textsubscript{c}$ and RF performance.

\subsection{Surface Morphology and Compositional Analysis of the Cu-Based Nb$_3$Sn Small Sample}

To assess the microstructural quality and phase composition of the Cu-based Nb$_3$Sn coating, scanning electron microscopy (SEM), energy-dispersive X-ray spectroscopy (EDX), and X-ray diffraction (XRD) were performed on a small witness sample prepared under identical conditions.

\begin{figure}[htbp]
  \centering
  \includegraphics[width=1.0\textwidth]{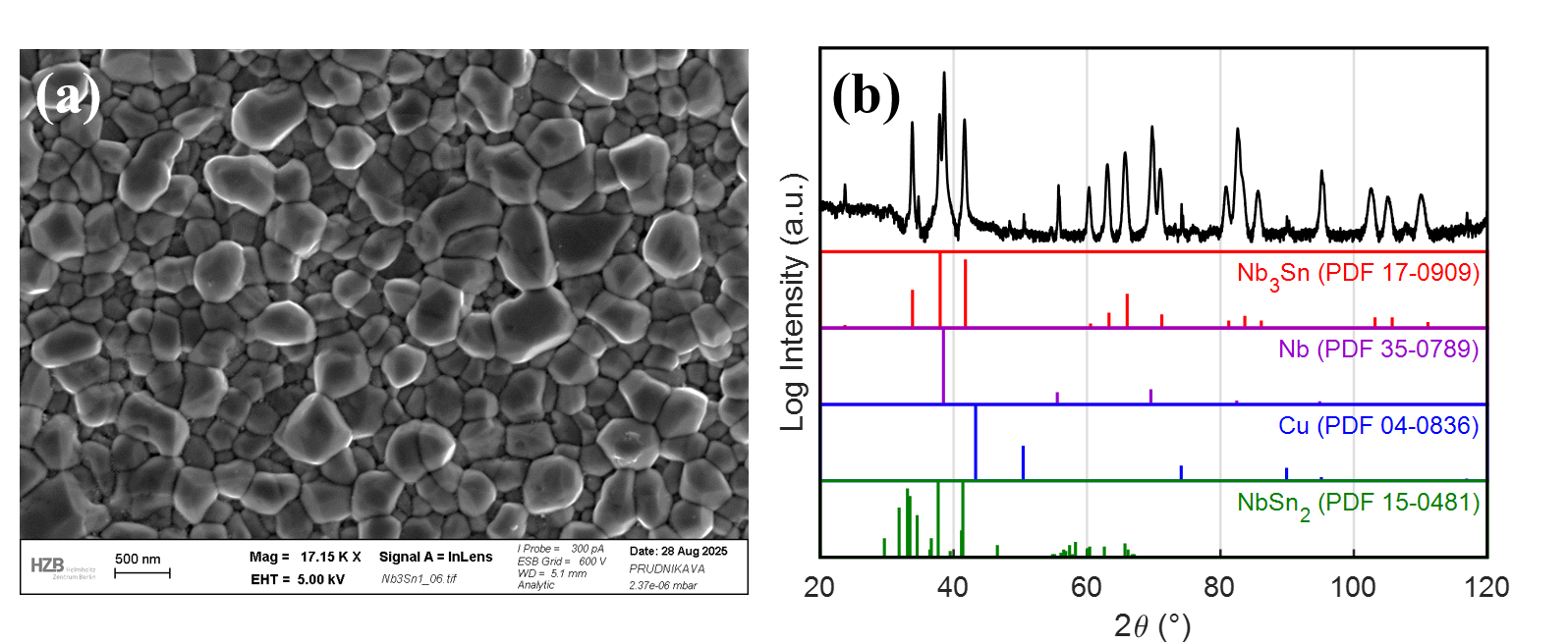}
  \caption{Surface characterization of the Cu-based Nb$_3$Sn sample. (a) SEM image showing a granular Nb\textsubscript{3}Sn surface with moderately fine grains and surface particulates. (b) XRD pattern indicating dominant A15 Nb\textsubscript{3}Sn phase along with weak contributions from secondary Nb–Sn intermetallics.}
  \label{fig:Nb3Sn_characterization}
\end{figure}

As shown in Fig.~\ref{fig:Nb3Sn_characterization}a, the SEM image reveals a continuous Nb\textsubscript{3}Sn coating consisting of granular crystallites with an average size of $\sim$180~nm. The grains appear well-connected, and no extended cracks or voids were detected. Nevertheless, the presence of surface impurity particulates and a broad Nb\textsubscript{3}Sn grain size distribution suggests local inhomogeneity, which may contribute to nonuniform RF losses. The XRD pattern in Fig.~\ref{fig:Nb3Sn_characterization}b, measured in Bragg--Brentano (BB) geometry, confirms that the dominant phase is the A15 Nb\textsubscript{3}Sn structure, with strong reflections such as (200), (210), and (211), consistent with polycrystalline A15 growth. The average crystallite size estimated from the (211) peak broadening is $\sim$150~nm, in good agreement with SEM. Besides the dominant A15 reflections, weak Nb, Cu, and NbSn\textsubscript{2} peaks are detected. The Nb signal originates from the diffusion barrier layer, while the Cu and NbSn\textsubscript{2} phases represent residual impurities that require further removal to ensure optimal SRF performance.

\begin{figure}[htbp]
  \centering
  \includegraphics[width=1.0\textwidth]{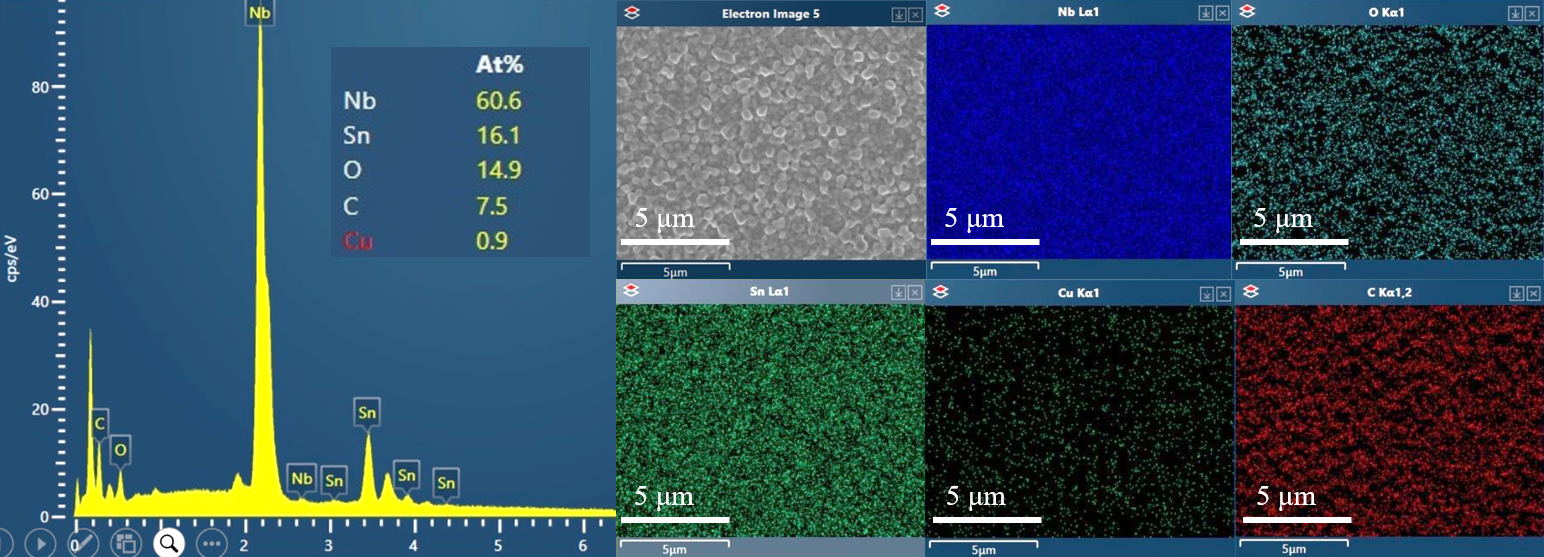}
  \caption{EDX analysis of the Cu-based Nb$_3$Sn sample, showing elemental composition and surface distribution.}
  \label{fig:Nb3Sn_EDX}
\end{figure}

The compositional analysis by EDX (Fig.~\ref{fig:Nb3Sn_EDX}) shows an average composition of Nb 60.6 at.\%, Sn 16.1 at.\%, O 14.9 at.\%, C 7.5 at.\%, and Cu 0.9 at.\%. Elemental mapping demonstrates that Nb and Sn are uniformly distributed across the surface, confirming the formation of a continuous Nb\textsubscript{3}Sn layer. However, the Sn content is below the stoichiometric 25 at.\%, which is partly attributed to Nb barrier layer contributions. The relatively high O and C fractions point to contamination, likely from surface oxidation and adsorbates, which are known to increase RF surface resistance in SRF applications. Although Cu is present only at trace levels, its normal-conducting nature can still contribute to RF surface resistance at low temperatures, making its further removal essential for achieving low-loss A15 films in SRF applications.

\begin{figure}[htbp]
    \centering
    \includegraphics[width=1.0\textwidth]{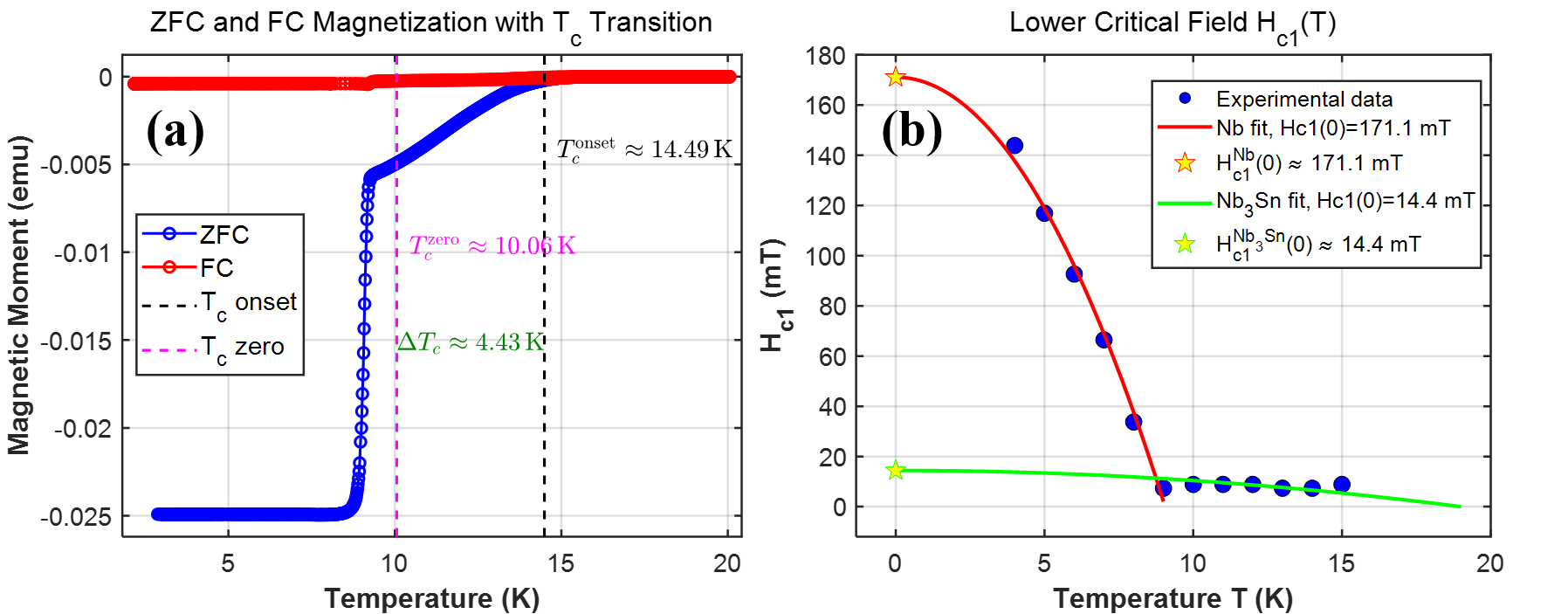}
    \caption{Magnetization and lower critical field measurements of the Cu-based Nb$_3$Sn sample. (a) ZFC/FC magnetization curves showing the superconducting transitions of the Nb barrier and Nb$_3$Sn layer. (b) Temperature dependence of the lower critical field $H_{c1}$ with empirical fitting results.}
    \label{fig:PPMS_results}
\end{figure}

The superconducting properties of the Cu-based Nb$_3$Sn sample were further investigated using a Physical Property Measurement System (PPMS), as shown in Fig.~\ref{fig:PPMS_results}. In Fig.~\ref{fig:PPMS_results}a, the zero-field-cooled (ZFC) and field-cooled (FC) magnetization curves reveal a superconducting transition with an onset temperature of $T_c^{\rm onset} \approx 14.49~\mathrm{K}$ and a zero-resistance temperature of $T_c^{\rm zero} \approx 10.06~\mathrm{K}$, corresponding to a transition width of $\Delta T_c \approx 4.43~\mathrm{K}$. The relatively broad transition suggests compositional and structural inhomogeneity, consistent with the SEM and XRD results. Fig.~\ref{fig:PPMS_results}b presents the temperature dependence of the lower critical field $H_{c1}(T)$ for both the Nb barrier and the Nb$_3$Sn layer. Fitting the data using $H_{c1}(T) = H_{c1}(0)\left[1-(T/T_c)^2\right]$ yields $H_{c1}^{\rm Nb}(0) \approx 171.1~\mathrm{mT}$ for the Nb layer and $H_{c1}^{\rm Nb_3Sn}(0) \approx 14.4~\mathrm{mT}$ for the Nb$_3$Sn coating. While the Nb film exhibits values consistent with bulk Nb, the reduced $H_{c1}(0)$ for Nb$_3$Sn reflects limited phase purity and possible weak-link effects at grain boundaries. 

From the perspective of SRF applications, these results demonstrate the successful formation of a superconducting Nb$_3$Sn layer on the Cu substrate. However, further optimization is required to improve stoichiometry, reduce Cu inclusions and tin-rich impurity phases, mitigate oxygen contamination, increase Nb$_3$Sn film density, and relieve its internal stress. Enlarging the grain size to minimize flux pinning and sharpening the superconducting transition are also essential. Such improvements are expected to raise the effective $T_c$ above 17~K and reduce the RF surface resistance, thereby enhancing the Nb$_3$Sn/Cu system for high-performance SRF applications.

\section{Conclusion and Outlook}

In this study, we have successfully demonstrated the first fabrication and RF characterization of a Cu-based Nb\textsubscript{3}Sn sample via an improved electrochemical–thermal synthesis (ETS) using the bronze route. The process integrates a series of critical advancements tailored for application on Cu substrates, including high-quality electrochemical polishing of the Cu surface, dense Nb diffusion barrier deposition via HiPIMS, uniform bronze precursor plating, controlled low-temperature annealing at 700\,\textdegree{}C, and an optimized chemical etching procedure for surface purification. These developments effectively address key material challenges such as Cu and Nb\textsubscript{3}Sn interdiffusion, compositional non-uniformity, and the formation of impurity phases. The resulting Nb\textsubscript{3}Sn coating achieved continuous coverage with clear A15 phase formation and minimal surface contamination. 

RF measurements using a multi-frequency QPR revealed promising performance of the Cu-based Nb\textsubscript{3}Sn sample. A minimum surface resistance of 43.4\,n$\Omega$ was achieved at 4.5\,K and 15\,mT at 413\,MHz, significantly outperforming earlier Cu-based Nb\textsubscript{3}Sn films synthesized via DCMS or HiPIMS + annealing. The superconducting transition temperature was measured to be approximately 14.6\,K, suggesting successful Nb\textsubscript{3}Sn phase formation, although the relatively broad transition indicates some degree of compositional inhomogeneity, likely due to Sn deficiency in certain regions. The frequency-dependent surface resistance and field-induced Q-slope observed at both 413\,MHz and 845\,MHz point toward extrinsic loss mechanisms, such as residual oxides, substoichiometric regions, and Cu inclusions. The extrapolated intrinsic quench field of 56.3\,mT, while encouraging, remains below the theoretical limit, emphasizing the need for further refinement in barrier layer design and thermal processing to suppress weak links and enhance grain boundary connectivity.

To further improve the performance of Nb\textsubscript{3}Sn coatings on Cu substrates via the ETS route, future efforts will focus primarily on optimizing the diffusion barrier design. This includes increasing the Nb barrier layer thickness to 30\,$\mathrm{\mu m}$ or more and enhancing its density to more effectively suppress Cu diffusion during annealing. The incorporation of an additional Ta diffusion barrier is also being considered. Concurrently, further refinement of the thermal treatment profile will be pursued to promote phase purity and uniform Sn distribution throughout the coating. Development of tailored chemical polishing and surface cleaning protocols for the Nb\textsubscript{3}Sn film will also be essential to reduce contamination and eliminate surface-related defects. These improvements aim to enable reliable, scalable fabrication of high-performance Nb\textsubscript{3}Sn/Cu structures for SRF applications.

Overall, this work provides compelling evidence for the feasibility of the bronze route in producing high-quality Nb\textsubscript{3}Sn coatings directly on Cu substrates at temperatures compatible with large-scale SRF cavity production. Compared to traditional Nb/Cu multilayers and bulk Nb structures, the Nb\textsubscript{3}Sn/Cu combination offers improved performance at elevated temperatures (e.g., 4.2–4.5\,K), reduced cryogenic load, and greater cost-efficiency. The current results represent an important milestone toward realizing fully functional Nb\textsubscript{3}Sn/Cu SRF cavities. Future efforts will focus on eliminating residual impurities, optimizing phase stoichiometry, and scaling the process to elliptical cavities operating at 1.3\,GHz. With continued process development, the electrochemical–thermal bronze route holds strong potential for enabling the next generation of high-performance, energy-efficient SRF accelerators.

\section*{Acknowledgements}

We gratefully acknowledge I.~Rudolph for granting access to the chemical facilities at the WI-APG department of HZB. We also thank R.~Feyerherm from the QM-IQM department at HZB for performing the PPMS measurements. This work has received partial funding from the European Union’s Horizon 2020 Research and Innovation programme under Grant Agreement No.~101004730 (I.FAST). 
Additional support was provided by the German Federal Ministry of Education and Research (BMBF) under Grant No.~100636506 (CavitySusOp, within the SuperSurfer framework).

\clearpage
\bibliographystyle{unsrt}

\end{document}